\documentclass[twocolumn,english,prl,showpacs,superscriptaddress,longbibliography]{revtex4-1}
\usepackage{amsfonts}
\usepackage{amsmath}
\usepackage{amssymb}
\usepackage[colorlinks, citecolor=blue,linkcolor=red]{hyperref}
\usepackage{color}
\usepackage{float}
\usepackage{hyperref}
\usepackage{textcomp}
\usepackage[rightcaption]{sidecap}
\usepackage{subfigure}
\usepackage[rightcaption]{sidecap}
\usepackage{graphicx}
\begin{document}
\title{Topological Nonlinear Optics with Spin-Orbit coupled Bose-Einstein Condensate in Cavity}
\author{Kashif Ammar Yasir}
\email{kayasir@iphy.ac.cn}\affiliation{Beijing National Laboratory for Condensed Matter Physics, Institute of Physics, Chinese Academy of Sciences, Beijing 100190, China.}
\affiliation{School of Physical Sciences, University of Chinese Academy of Sciences, Beijing 100190, China.}
\author{Lin Zhuang}
\email{stszhl@mail.sysu.edu.cn}\affiliation{School of Physics, Sun Yat-Sen University, Guangzhou 510275, China.}
\author{Wu-Ming Liu}
\email{wliu@iphy.ac.cn}\affiliation{Beijing National Laboratory for Condensed Matter Physics, Institute of Physics, Chinese Academy of Sciences, Beijing 100190, China.}
\affiliation{School of Physical Sciences, University of Chinese Academy of Sciences, Beijing 100190, China.}
\setlength{\parskip}{0pt}
\setlength{\belowcaptionskip}{-10pt}
\begin{abstract}
We report topological nonlinear optics with spin-orbit coupled Bose-Einstein condensate in a cavity. The cavity is driven by a pump laser and weak probe laser which excite Bose-Einstein condensate to an intermediate storage level, where the standard Raman process engineers spin-orbit coupling. 
We show that the nonlinear photonic interactions at the transitional pathways of dressed states result in new type of optical transparencies, which get completely inverted with atom induced gain.
These nonlinear interactions also implant topological sort of features in probe transmission modes by inducing gapless Dirac-like cones, which become gaped in presence of Raman detuning.
The topological features get interestingly enhanced in gain regime where the gapless topological edge-like states emerge among the probe modes, which can cause non-trivial phase transition.      
We show that spin-orbit coupling and Zeeman field effects also impressively revamp fast and slow probe light. The manipulation of dressed states for quantum nonlinear optics with topological characteristics in our findings could be a crucial step towards topological quantum computation.
\end{abstract}
\date{\today}
\maketitle

The quantum nonlinear optics, a stunning manipulation of nonlinear and strong photonic interactions at single photon level \cite{Ref1,Ref2,Ref3}, has the ability to practically demonstrate quantum computation \cite{Ref4,Ref5,Ref6}. The electromagnetically induced transparency (EIT) -- quantum interference between two photon transitional paths excited from a single state \cite{Ref7} -- is the finest example of such nonlinear interactions \cite{Ref8,Ref9}. These quantum interferences completely stop or slow-down \cite{Ref18} the probe light during the EIT interval which impressively appeals for the optical storage devices \cite{Ref4,Ref7,Ref91}.
The coherent environment of optical cavities significantly enhances these optical nonlinear phenomena \cite{Ref10}, which yield in various significant advancements \cite{Ref11,Ref12,Ref13,Ref14,Ref15,Ref16,Ref30}. Further, the mechanical effects of light \cite{Ref18,Ref17,Ref19,Ref20} lead to optomechanically induced transparency (OMIT) \cite{Ref21,Ref22,Ref23} which, in coupling with other objects (like Bose-Einstein condensate (BEC)), result in multiple transparencies \cite{Ref24,Ref25,Ref26,Ref27,Ref28,Ref29} in a single cavity. However, it's worth wondering that how nonlinear photonic interactions are significant to approach the topological defects in light. 

The revelation of topological states of light has emerged as a milestone in quantum optics \cite{Ref53,Ref54,Ref55,Ref56,Ref57,Ref59,Ref60}. The topological characteristics of photons have been studied in gyromagnetic photonic crystals \cite{Ref571}, quasicrystals \cite{Ref572}, optical waveguides \cite{Ref573} and many other systems \cite{Ref53}. Here the behavior of parity-time (PT)-symmetry plays a crucial role for quantum phase transition in photons \cite{Ref53,Refa0,Refa1,Refa2,Refa3} and has led to many stunning phenomena, like loss-induced transparency \cite{Refa4}, low power optical diodes \cite{Refa5} and single-mode lasers \cite{Refa6,Refa7}. Furthermore, the discovery of broken optical PT-symmetry in two coupled optomechanical oscillators \cite{Refa8,Refa9,Refa10,Refa11,Refa12}, by considering one in gain and other in lossy regime, has led to the demonstration of low threshold optical chaos \cite{Refa13}, inverted OMIT \cite{Refa14} and phonon laser \cite{Refa15}.
However, to utilize topological light for quantum information, the manipulation of topological phases of light with nonlinear optics is highly desirable, which we intend to achieve with spin-orbit (SO)-coupled dressed states \cite{Ref31,Ref32,Ref33,Ref34} in a single cavity.
As the SO-coupling is essential to study spin-Hall effect \cite{Ref35,Ref36} and topological insulators \cite{Ref37,Ref38,Ref39,Ref40,Ref41,Ref42,Ref43,Ref44,Ref441,Ref442,Ref443,Ref444,Ref445,Ref446,Ref447,Ref448} with atoms, so it is worthwhile to see its effects on transmitted light. 
The inclusion of SO-coupled BEC in optical (and optomechanical) cavities \cite{Ref45,Ref46,Ref47,Ref48} and the observation of topological edge state \cite{Ref58} as well as Weyl symmetries \cite{Ref61} in cavity transmission have further motivated us to use synthetic SO-coupling for topological-nonlinear optics.

In this Letter, we demonstrate topological characteristics of nonlinear optics mediated by SO-coupled BEC in a cavity. The SO-coupled dressed states excite to the intermediate state after interacting with strong pump and weak probe field, where a first EIT window appears. As the Raman process splits the BEC into hyperfine states, which than become momentum sensitive \cite{Ref31,Ref32,Ref33,Ref34}, so the quantum interference at these states opens another opportunity for the dark-state. We not only show that the photon interactions during the probe transition through the dressed states and the Zeeman shift generate EIT-like transparencies but also illustrate that the nonlinear interaction of dressed states induces gapless and gaped Dirac-like modes in probe transmission, preserved by the PT-symmetry, depending upon Raman detuning. However, in amplification case -- where the atomic damping greater than cavity decay acts as a gain to the cavity field \cite{Refb} -- an edge-like topological mode appears between upper and lower band revealing the broken PT-symmetry. Furthermore, we show that the Zeeman field effects also dramatically alter the fast and slow dynamics of transmitted probe light.    
 
\begin{figure}[tp]
	\includegraphics[width=6.cm]{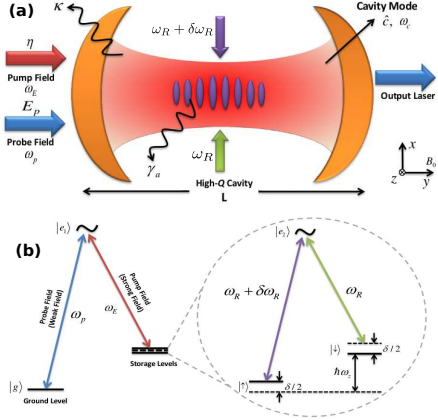}
	\caption{(a) Schematic setup for $^{87}Rb$ spin-orbit coupled Bose-Einstein condensate inside 
		a high-\textit{Q} cavity, driven by control pump field $\omega_E$ and a probe field $\omega_p$. The bias magnetic field $B_0$ is applied along the cavity ($\hat{y}$-axis) while the Raman beams interact transversally ($\hat{x}$-axis) with atoms. (b) The energy level excitation diagram, where left-side illustrates the probe-pump excitation of the atoms and right-side is the Raman excitations at storage level for spin-orbit coupling.}
	\label{fig11}
\end{figure}
The $^{87}Rb$ bosonic particles, with $N\approx1.8\times10^5$, are trapped inside a high-finesse optical cavity with length $L\approx1.25\times10^{-4}$m, Fig. \ref{fig11}(a). The external pump laser $\omega_{E}=\omega_R+\delta\omega_R=8.8\times2\pi$ MHz and a weak probe laser $\omega_p<<\omega_E$ both driven the cavity and excite atomic mode to an intermediate storage state with electronic levels $5S_{1/2}$ \cite{Ref48}. 
Because of the high-\textit{Q} factor, external pump field builds a strong cavity mode oscillating at $\omega_{c}=4\times2\pi MHz$ with detuning $\Delta_c=\omega_E-\omega_c=\delta\omega_R$.
The $10 G$ bias longitudinal magnetic field $B_0$ produces a Zeeman splitting $\hbar\omega_z$  with $\omega_z\approx4.8\times2\pi$ kHz.
To produce SO-coupling, two Raman transverse lasers ($\omega_R$ and $\omega_R+\delta\omega_R$), where $\delta\omega_R=\omega_z+\delta/\hbar\backsimeq4.8\times2\pi$ MHz with wavelength $\lambda=804.1 nm$ and detuning $\delta=1.6E_R$ \cite{Ref34}, oppositely interact with atoms along $\hat{x}$-axis. This Raman process for SO-coupling yields in the coupling of two internal pseudo-spin states ($|\uparrow\rangle =| F=1, m_F=0\rangle$ and $|\downarrow\rangle=| F=1, m_F=-1\rangle$) at same electronic manifold $F=1$, see Fig. \ref{fig11}(b).      

The complete system Hamiltonian, under rotating-wave approximation \cite{Ref34,Ref46,Ref47,Ref48}, can be expressed as,
\begin{eqnarray}\label{Ha1}
\hat{\mathcal{H}} &=& \int d\pmb{r}\pmb{\hat{\psi}^{\dag}}(\pmb{r})\bigg(\hat{\mathcal{H}}_{SOC}+\mathcal{V}\bigg) \pmb{\hat{\psi}}(\pmb{r})\nonumber\\
&+&\frac{1}{2}\int d\pmb{r}\sum_{\sigma,\acute{\sigma}} \mathcal{U}_{\sigma,\acute{\sigma}}\hat{\psi}^{\dag}_\sigma(\pmb{r})\hat{\psi}^{\dag}_{\acute{\sigma}}(\pmb{r})\hat{\psi}_{\acute{\sigma}}(\pmb{r})\hat{\psi}_\sigma(\pmb{r})+\hbar\triangle_{c}\hat{c}^{\dag}\hat{c}\nonumber\\
&-&i\hbar\eta(\hat{c}
-\hat{c}^{\dag})-i\hbar \mathcal{E}_p(\hat{c}e^{i\Delta_pt}-\hat{c}^\dagger e^{-i\Delta_pt}),
\end{eqnarray}
where $\pmb{\hat{\psi}}=[\hat{\psi}_\uparrow,\hat{\psi}_\downarrow]^{T}$ is the notation of bosonic field operators for atomic $|\uparrow\rangle$ and $|\downarrow\rangle$ states. 
$\hat{\mathcal{H}}_{SOC}=\hbar^2\pmb{k}^2\sigma_{0}/2m_a+\tilde{\alpha}\pmb{k_x}\sigma_{y}+\frac{\delta}{2}\sigma_y+\frac{\Omega_z}{2}\sigma_z$ 
is the Hamiltonian for a SO-coupled single particle, containing the strength of SO-coupling $\tilde{\alpha}=E_R/\pmb{k_L}$ and the magnetic field effects $\delta=-g\mu_B B_z$ (Raman detuning) and $\Omega_z=-g\mu_B B_y$ (Raman coupling) along the $\hat{z}$ and $\hat{y}$ axis, respectively \cite{Ref34,Ref48}. The quasi-momentum will be one-dimensional $\pmb{k}=[\pmb{k_x},0,0]$ as the SO-coupling only occurs along $\hat{x}$-axis. $\sigma_{x,y,z}$ are the $2\times2$ Pauli matrices with unit matrix $\sigma_0$.
$\mathcal{V}=\hbar \hat{c}^{\dag}\hat{c}U_0[cos^2(kx)+cos^2(ky)]$ is the two-dimensional optical lattice formed by the longitudinal and transverse fields \cite{Ref48}, where $\hat{c}$ ($\hat{c}^\dag$) is an annihilation (creation) operator for the cavity field. $U_0=g_0^2/\Delta_a$ is the optical depth for the atoms defined by the Rabi oscillation $g_{0}$ and atomic-cavity detuning $\Delta_{a}$ \cite{Ref48}. The strength of atom-atom interactions is defined by $\mathcal{U}_{\sigma,\acute{\sigma}}=4\pi a_{\sigma,\acute{\sigma}}^2\hbar^2/m_a$, where $a_{\sigma,\acute{\sigma}}$ accounts for s-wave scattering. Last two terms define the relation of cavity field with external pump and probe field, with intensities  $\vert\eta\vert=\sqrt{P\kappa/\hbar\omega_{E}}$ and $\vert\eta_p\vert=\sqrt{P_p\kappa/\hbar\omega_{p}}$, respectively.  

We consider the strength of intra-species and inter-species interactions as $U_{\uparrow,\uparrow}=U_{\downarrow,\downarrow}=U$ and $U_{\uparrow,\downarrow}=U_{\downarrow,\uparrow}=\varepsilon U$, respectively, with laser configuration factor $\varepsilon$ \cite{Ref34,Ref48}.
By applying plane-wave approximation for bosonic wave-function $\pmb{\hat{\psi}}(r)=e^{ikr}\pmb{\hat{\varphi}}$, with $\pmb{\hat{\varphi}}=[\hat{\varphi}_\uparrow,\hat{\varphi}_\downarrow]^{T}$, under normalization condition 
$|\hat{\varphi}_\uparrow|^2+|\hat{\varphi}_\downarrow|^2=N$, we derive the coupled Langevin equation equations for the system \cite{Ref48,Ref49}. We linearized Langevin equations to incorporate first order quantum fluctuations over their steady-state values $\mathcal{\hat{O}}(t)=\mathcal{O}_{s}+\mathcal{\delta O}(t)$, here $\mathcal{O}$ and $\mathcal{O}_{s}$ correspond to the associated subsystems and their steady-states, respectively. By considering equal amount of particles in each species under normalization conditions, which yield in $\hat{\varphi_\uparrow}^\dag\hat{\varphi_\uparrow}=\hat{\varphi_\downarrow}^\dag\hat{\varphi_\downarrow}=N/2$, we define dimensionless position and momentum quadratures for atomic subsystems, $\hat{q}_\mathcal{N}=\frac{1}{\sqrt{2}}(\hat{\mathcal{O}}+\hat{\mathcal{O}}^{\dag})$ and $\hat{p}_\mathcal{N}=\frac{i}{\sqrt{2}}(\hat{\mathcal{O}}-\hat{\mathcal{O}}^{\dag})$, respectively, where $\mathcal{N}\ni\{\uparrow,\downarrow\}$. Under this configuration, the linearized equations will read as,
\begin{eqnarray}
\partial_t\delta\hat{c}(t)=-&(\kappa&+i\Delta)\delta\hat{c}(t)-iG(\delta\hat{q}_\uparrow(t)+\delta\hat{q}_\downarrow(t))
+\eta_pe^{-i\Delta_pt}\label{eq1},\nonumber\\
\partial_t\delta\hat{p}_{\uparrow,\downarrow}(t)&=&\mathcal{M}\delta\hat{p}_{\uparrow,\downarrow}(t)\pm\frac{\Omega_z}{2}\delta\hat{q}_{\uparrow,\downarrow}(t)\mp(\alpha-\frac{\delta}{2})\delta\hat{p}_{\downarrow,\uparrow}(t)\nonumber\\
&&-\sqrt{2}G(\delta\hat{c}(t)+\delta\hat{c}^\dagger(t)),\\
\partial_t\delta\hat{q}_{\uparrow,\downarrow}(t)&=&\mathcal{M}\delta\hat{q}_{\uparrow,\downarrow}(t)\pm\frac{\Omega_z}{2}\delta\hat{p}_{\uparrow,\downarrow}(t)\pm(\alpha-\frac{\delta}{2})\delta\hat{q}_{\downarrow,\uparrow}(t).\nonumber
\end{eqnarray}

The $\partial_t$ corresponds to the first order time derivative. The parameter $M=\Omega/2+v-\gamma+UN(1-\epsilon)$, where $v=g_an_s$ and $\Omega=\hbar\pmb{k}^2/m_a$, while $G=\sqrt{2}g_{a}|c_{s}|$ and $\Delta = \Delta_{a} +g_{a}N$ are the effective atom-field coupling defined over $g_a$, and detuning of the system, respectively. The collective density excitations of each atomic species perform oscillations under cavity radiation pressure, mimicking as two atomic mirrors equally coupled to the cavity field. $g_{a}=\frac{\omega_{c}}{L}\sqrt{\hbar/m_{bec}\Omega}$ defines their optomechanical coupling strength, with mass $m_{bec}=\hslash\omega_{c}^{2}/(L^{2}U^2_{0}\Omega)$ \cite{Ref48,Ref49}. In order to determine and develop sufficient stability conditions, we use Routh-Hurwitz Stability Criterion \cite{Ref29,Ref48,Ref49} and strictly follow these conditions in numerical calculations.  

\begin{figure}[tp]
	\includegraphics[width=8.5cm]{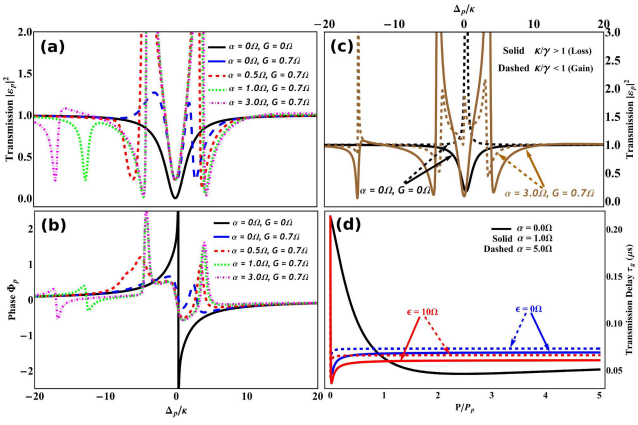}
	\caption{(a) The transmission amplitude $|\mathcal{E}_p|^2$ and (b) the transmission phase $\Phi_p$ verses $\Delta_p/\kappa$, for various values of $G/\Omega$ and $\alpha/\Omega$ at $\varepsilon/\kappa=1$. (c) $|\mathcal{E}_p|^2$ verses $\Delta_p/\kappa$ in lossy $\kappa/\gamma>1$ (solid) and gain $\kappa/\gamma<1$ (dashed) regimes with absence (black) and presence (brown) of SO-coupled BEC. (d) The transmission delay $\tau_g$ verses pump power $P/P_p$ illustrating the influences of $\epsilon/\Omega$ and $\alpha/\Omega$. The other parameters used are $\Omega_z=\Omega$, $\delta=0$, $U/\Omega=5.5$, $\Delta=1.5\kappa$ and $\Omega=3.8\times2\pi$MHz.}
	\label{fig12}
\end{figure}
We obtain cavity transmission by solving linearized Langevin equations, for details see Ref.\cite{Ref491}. The total probe transmission, from atom-cavity, can be obtained by taking the ratio of probe input and output probe components $\mathcal{E}_p(\omega_p)=(\eta_p-\sqrt{2k}c_-(\Delta_p))/\eta_p=1-\sqrt{2k}c_-(\Delta_p)/\eta_p$. Here amplitude $\varepsilon_{out}=\frac{\sqrt{2k}c_-(\Delta_p)}{\eta_p}$, at $\omega_p$ and $\Delta_p=\omega_e-\omega_p$, determines the absorption (in-phase) and dispersion (out-of-phase) cavity transmission with its real and imaginary values. The transmission phase, i.e. $\Phi_p(\omega_p)=arg(\mathcal{E}_p)$ is crucially important to understand whole picture because, with its rapid dispersion, the dynamics of fast and slow probe transmission can be governed by computing transmission (or group) delay $\tau_g=\partial\Phi_p(\omega_p)/\partial\omega_p$, as illustrated in Fig. \ref{fig12}.  

In the empty cavity case, the weak probe field will simply pass the cavity without experiencing any quantum interference, as illustrated by the black curves in Fig. \ref{fig12}(a) and \ref{fig12}(b). However, when external fields resonantly interacts with intra-cavity atoms, they excite them to the intermediate energy level opening a way for quantum interference to generate the first EIT, as illustrated in the blue curves of $|\mathcal{E}_p|^2$ and $\Phi_p$ of Fig. \ref{fig12}, at $\Delta_p\approx2.5\kappa$. This type of EITs have been extensively studied for the cavity containing cold/ultra-cold atoms \cite{Ref7,Ref20,Ref26,Ref27,Ref28,Ref29}, even for the single atom and ion \cite{Ref11,Ref12}.
\begin{figure}[tp]
	\includegraphics[width=9cm]{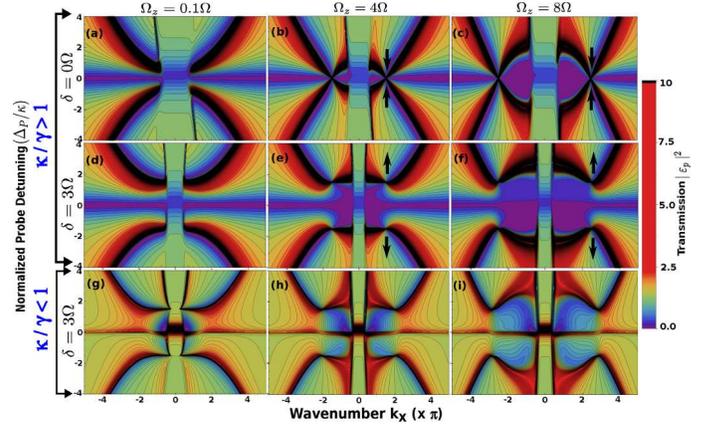}
	\caption{The transmission amplitude $|\mathcal{E}_p|^2$ verses $\mathbf{k_x}(\times\pi)$ and $\Delta_p/\kappa$. The first, second and third rows illustrate the influence of $\Omega_z/\Omega$ at $\delta=0\Omega$, $3\Omega$ and $\delta=3\Omega$, respectively. The first and second rows correspond to the lossy regime ($\kappa/\gamma>1$) while the third row is for gain regime ($\kappa/\gamma<1$).}
	\label{fig13}
\end{figure}

The standard Raman process, in presence of magnetic field $B_0$, at the intermediate state further splits the atomic state into two hyperfine states, $|\uparrow\rangle$ and $|\downarrow\rangle$, by dressing their electronic manifolds, see Fig. \ref{fig11}(b). The SO-coupled atomic states acts as another double excitation forming at storage state, which eventually generates a second dark-state with quantum interference between transitional pathways of $|\uparrow\rangle$ and $|\downarrow\rangle$. These interactions result in new type of nonlinear photon interactions because of SO-coupling, as shown by the EIT window appearing in red curves of $|\mathcal{E}_p|^2$ and $\Phi_p$. Moreover, when we increase the SO-coupling, another interesting EIT-like structure appears in the transmission spectrum, as illustrated by green and magenta curves in Fig. \ref{fig12}(a) and \ref{fig12}(b), when $\alpha=1\Omega$ and $3\Omega$, respectively. The third EIT-like structure appears because at higher SO-coupling, the Zeeman shift, as shown by energy gap $\hbar\omega_z$ in Fig. \ref{fig11}(b), become more affective. These magnetic effects for photon interactions can be modeled as an analog of Magnetically induced transparency (MIT) \cite{Ref30}. The position of the third structure varies with applied $\alpha/\Omega$ but it can be adjust with $G/\Omega$ and $\epsilon$, see Ref. \cite{Ref491}. The inter-species interactions $\epsilon$ interestingly enhance the SO-coupling effects on probe transmission providing an agreement for increase in quantum interferences with SO-coupling. 

The dynamics of PT-symmetry plays an important role to describe quantum phase transition in photons \cite{Ref53,Refa0,Refa1} and crucially depends on the loss and gain of optical fields after interacting with passive and active medium \cite{Refa8,Refa9,Refa10,Refa11,Refa12,Refa13,Refa14,Refa15}, respectively. As, in our system, the dressed states acts as equally coupled oscillators inside the cavity which damp after interacting with cavity field, therefore, one can model these oscillators as active subsystems whose damping acts as gain for the probe transmission \cite{Refa14,Refb}. It is possible under condition that the cavity decay rate is smaller than the atomic damping rate $\kappa<\gamma$. In the gain configuration, the probe transmission, for both non-EIT as well as EIT case, get completely inverted \cite{Refa14} because of the amplification (gain) in probe field, see Fig. \ref{fig12}(c). In empty cavity case, $|\mathcal{E}_p|^2$ moves from absorption domain (i.e. $|\mathcal{E}_p|^2<1$) to amplification domain (i.e. $|\mathcal{E}_p|^2>1$) while for coupled cavity case EIT spectrum get inverted. Here it should be noted that, at $\kappa/\gamma>1$, the amplitudes of EIT-windows are partially appearing in amplification domain $|\mathcal{E}_p|^2>1$, and vise-versa, which may be the reason behind the weak topological transport between bulk modes in lossy regime, as discussed later.     
Further, the rapid phase diffusion with $\alpha$ also prolongs the probe passage from the dressed states, as shown in Fig. \ref{fig12}(d) where $\tau_g$ increases with the increase in $\alpha$. It also shows the slow and fast light dynamics can also be controlled with $\epsilon$.                

The interesting thing arrives when the synthetic process of SO-coupling leaves topological-like signatures to the probe field \cite{Ref53,Ref54,Ref58,Ref61}. These features can be visualized by plotting probe transmission components parallel to the direction of SO-coupling (i.e. $x$-axis), as illustrated in Fig. \ref{fig13}. As the probe field is appeared to be the function of $\Delta_p$ and $\mathbf{k_x}$ is parallel to the wavenumber of probe field along $x$-axis, therefore, the gaped and gapless modes can be obtained by plotting $|\mathcal{E}_p|^2$ verses $\mathbf{k_x}$ and $\Delta_p/\kappa$.
At weak Raman coupling $\Omega_z=0.1\Omega$, no Dirac cone type (bulk states) structure appears in the transmission field resulting in trivial transmission. The two side modes can be seen at $\Delta_p/\kappa>0$ and $\Delta_p/\kappa<0$ corresponding to the incoherent creation and annihilation of quasi-particles, respectively, by the probe field. However, when we increase $\Omega_z/\Omega$, two symmetrical Dirac cones take form at $\mathbf{k_x}\approx\pm2\pi$, as indicated with black arrows in Fig. \ref{fig13}(c) and (d). These Dirac comes result in a gap-less Weyl points oriented along the $x$-axis in $x-y$-plane \cite{Ref61}. The gap-less Dirac cones like features show agreement for the topological phase transitions of optical mode. It happens because the gap emerging between atomic spin-states \cite{Ref48} modifies the phase of probe field which then appears in new synthetic optical spin-states. 

These bulk states are crucially protected by the $PT$ symmetry \cite{Ref53} which, in our case, can be modified with the Raman detuning, as illustrated in Fig. \ref{fig13}(d-f) where the gap appears between Dirac cones with increase in $\delta/\Omega$ \cite{Ref58,Ref61}. It is obvious because, in presence of $\delta/\Omega$, the energy spectrum of the dressed states became asymmetric which eventually cause the separation (or broke the $P$ type symmetry \cite{Ref53}) of probe bulk and edge-like modes. A weak transport between bulk modes can be seen around $\mathbf{k_x}\approx0$, because of the amplified components of the probe field in lossy regime, but this can be significantly enhanced by considering optical gain regime $\kappa/\gamma<1$. A topological transport between edge-like state appears at $\mathbf{k_x}\approx0$ and $\mathbf{k_x}\approx\pm2\pi$ (illustrated by the increase in strength of transmission field) which closes the gap between top and bottom interface (optical spin-states), as illustrated in Fig. \ref{fig13}(g-i). The reason behind the emergence of these edge modes can be the broken PT-symmetry, particularly T-symmetry. As discussed earlier, in the gain regime ($\kappa/\gamma<1$), the probe amplitude in output get categorically shifted from absorption to amplification domain interrupting PT-symmetry. The gapless edge modes trigger robust transport for non-trivial optical phase transition, especially to the quantum Hall phase with non-zero Chern number.

\begin{figure}[tp]
	\includegraphics[width=8cm]{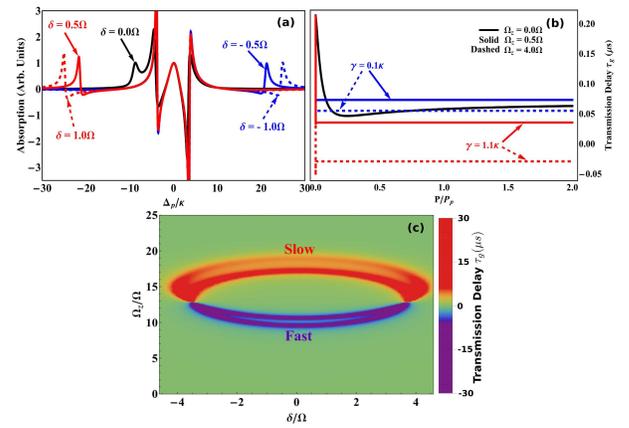}
	\caption{(a) The absorption spectrum $Re[\varepsilon_{out}]$ verses $\Delta_p/\kappa$ under influence of $\delta/\Omega$. (b) The transmission delay $\tau_g$ verses $P/P_p$ under influence of $\Omega_z/\Omega$ in lossy $\kappa/\gamma>1$ and gain $\kappa/\gamma<1$ regime. (c) The slow and fast dynamics of probe light (with $\tau_g$) as a function of $\Omega_z/\Omega$ and $\delta/\Omega$ when $P=1.5P_p$. $\alpha=2\Omega$ in these results.}
	\label{fig14}
\end{figure}

As $\alpha$ and $\Omega_z$ are interconnected in Raman process, so $\Omega_z$ will induce similar features in the probe transmission, as discussed in Ref. \cite{Ref491}. However, $\delta$ significantly effects the position and size of third transparency window, as illustrated by the blue and red curves of Fig. \ref{fig14}(a). The presence of $\delta$ alters the size of Zeeman gap which consequently appears as an asymmetric variations in eigen-energies spectrum of dressed states \cite{Ref48}. As the sign of $\delta$ is responsible for the $\pm\mathbf{k_x}$ direction of the asymmetric potential, therefore, it similarly shifts the third transparency over $\pm\Delta_p$, as can be seen in Fig. \ref{fig14}(a). The $\Omega_z$ also generates multiple Fano line-shapes in probe transmission which are significantly important for optical spectroscopy \cite{Ref50,Ref51}, for details see Ref. \cite{Ref491}. Furthermore, the increase in $\Omega_z$ is appear to be increasing the speed of probe light but the lossy ($\kappa/\gamma>1$) and gain ($\kappa/\gamma<1$) regime act as a switch between slow and fast probe transmission, as shown in Fig. \ref{fig14}(b), where $\gamma>\kappa$ significantly decreases $\tau_g$ to the fast light regime at certain $\Omega_z$. The behavior of slow and fast probe follow an interesting pattern against $\Omega_z$ and $\delta$, see Fig. \ref{fig14}(c), where both slow and fast light follow a semi-circular shape around $\Omega_z\approx13\Omega$, $\Omega_z>13\Omega$ slow and $\Omega_z<13\Omega$ fast, over a specific interval of $\delta$. These semi-circles of slow and fast transmission dramatically meet each other $\delta\approx\pm3.5\Omega$, at $\Omega_z\approx13\Omega$, which is also the point where maximum slow as well as fast light occurs. However, these dynamics can be altered with the help of other parameters as discussed in Ref. \cite{Ref491}.                

In conclusion, we propose a scheme to engineer topological nonlinear optics with SO-coupled BEC in an optical cavity. The quantum interferences during the probe transitions at dressed states and at Zeeman gap not only give birth to a new type of nonlinear optical transparencies which can be inverted in gain regime, but also imprint the gapless and gaped Dirac (bulk) modes in probe transmission, revealing the symmetric Weyl-like modes in gapless regime. We also illustrate the emergence of gapless edge-like topological modes in gain regime uncovering the broken PT-symmetry for non-trivial photonic phase transitions. Furthermore, we investigate the dynamics of fast and slow probe light under influence of dressed states and show that the speed of probe light can be increase or decrease with SO-coupling and Raman coupling. Interestingly, lossy and gain regimes act as a switch between the slow and fast light, respectively. Our findings provide new direction for quantum nonlinear optics with topological features, mediated by SO-coupled states, which could be crucial in order to bring topological physics to the quantum information science.

\begin{acknowledgments}
This work was supported by the NKRDP under grant No. 2016YFA0301500, NSFC under grant Nos. 11434015, 61227902 and SPRPCAS under grant Nos. XDB01020300, XDB21030300.
We also acknowledge the financial support from CAS-TWAS president's fellowship programme (2014).
\end{acknowledgments}

\end{document}


\title{Supplemental Material : Topological Nonlinear Optics with Spin-Orbit coupled Bose-Einstein Condensate in Cavity}
\author{Kashif Ammar Yasir}
\email{kayasir@iphy.ac.cn}\affiliation{Beijing National Laboratory for Condensed Matter Physics, Institute of Physics, Chinese Academy of Sciences, Beijing 100190, China.}
\affiliation{School of Physical Sciences, University of Chinese Academy of Sciences, Beijing 100190, China.}
\author{Lin Zhuang}
\email{stszhl@mail.sysu.edu.cn}\affiliation{School of Physics, Sun Yat-Sen University, Guangzhou 510275, China.}
\author{Wu-Ming Liu}
\email{wliu@iphy.ac.cn}\affiliation{Beijing National Laboratory for Condensed Matter Physics, Institute of Physics, Chinese Academy of Sciences, Beijing 100190, China.}
\affiliation{School of Physical Sciences, University of Chinese Academy of Sciences, Beijing 100190, China.}

\date{\today}
\maketitle

\subsection{Cavity output field calculations}
After considering the coupling of external pump field with cavity field larger than the probe field coupling (i.e. $\eta>>\mathcal{E}_p$), we substitute system quadratures with $\delta\mathcal{B}=\sum_{n\rightarrow\{+,-\}}\mathcal{B}_ne^{in\omega t}$, where $\mathcal{B}$ can be any associated subsystem. Here it should be noted that we have not incorporated the influences of associated quantum noises which is possible when the atomic mirrors oscillate at high frequency $\hbar\Omega>>k_BT$. In order to extract the components of probe field from linearized Langevin equations, we compare the coefficient of exponentials terms, which appear to be with exponent of probe detuning $\Delta_p=\omega_E-\omega_p$, and then solve them for intra-cavity field. To calculate output optical spectrum, we use standard input-output relation $c_{out}=\sqrt{2\kappa}c-c_{in}$ and extend the output field in the following expression,
\begin{eqnarray} 
\langle c_{out}\rangle=&&\eta-\sqrt{2k}c_s+(\eta_p-\sqrt{2k}c_-(\Delta_p))\mathcal{E}_pe^{-i\Delta_pt}\nonumber\\
&&-\sqrt{2k}c_+(\Delta_p)e^{i\Delta_pt}.
\end{eqnarray}

Here, as we know, the second term accommodates the probe field component in the cavity output and by comparing input-output fields (using input-output relation), one can obtain the expression for $c_{-}(\Delta_p)$,
\begin{eqnarray}
c_{-}(\Delta_p)=\frac{\eta_p(1-\mathcal{X}_{\uparrow\uparrow}(\Delta_p))}{( i\Delta+\kappa- i\Delta_p)(-1+\mathcal{X}_{\uparrow\uparrow}(\Delta_p)+\mathcal{X}_{\downarrow\downarrow}(\Delta_p))},
\end{eqnarray}
where,
\begin{eqnarray}
\mathcal{X}_{\uparrow\uparrow}(\Delta_p)&=&\frac{i2\sqrt{2}\mathcal{X}_\uparrow G^2}{ -i\Delta+\kappa-i\Delta_p},\\
\mathcal{X}_{\downarrow\downarrow}(\Delta_p)&=&\frac{i2\sqrt{2}\mathcal{X}_\downarrow G^2}{ i\Delta+\kappa-i\Delta_p}.
\end{eqnarray}
 $\mathcal{X}_{\uparrow\uparrow}(\Delta_p)$ and $\mathcal{X}_{\downarrow\downarrow}(\Delta_p)$ can be refereed as modified susceptibilities for atomic state $|\uparrow\rangle$ and $|\downarrow\rangle$, respectively, in the presence of spin-orbit (SO)-coupling, where
\begin{widetext}
 \begin{eqnarray}
 \mathcal{X}_{\uparrow,\downarrow}(\Delta_p)=&&\frac{(\alpha-\frac{\delta}{2}\mp i\Delta_p-\mathcal{M})\bigg(\pm(\mathcal{L}_{\uparrow,\downarrow}(\Delta_p)-\Omega_z^2)(i\Delta_p\mp\mathcal{M})+(\alpha-\frac{\delta}{2})(\mathcal{L}_{\uparrow,\downarrow}(\Delta_p)+\Omega_z^2)\bigg)}{(\mathcal{L}_{\uparrow,\downarrow}(\Delta_p)-\Omega_z^2)^2(i\Delta_p\mp\mathcal{M})^2+(\alpha-\frac{\delta}{2})^2(\mathcal{L}_{\uparrow,\downarrow}(\Delta_p)\pm\Omega_z^2)}\nonumber\\
 \pm&&\frac{(\alpha-\frac{\delta}{2}\mp i\Delta_p+\mathcal{M})\bigg((\mathcal{L}_{\uparrow,\downarrow}(\Delta_p)-\Omega_z^2)(i\Delta_p\mp\mathcal{M})\mp(\alpha-\frac{\delta}{2})(\mathcal{L}_{\uparrow,\downarrow}(\Delta_p)+\Omega_z^2)\bigg) }{(\mathcal{L}_{\uparrow,\downarrow}(\Delta_p)-\Omega_z^2)^2(i\Delta_p\mp\mathcal{M})^2+(\alpha-\frac{\delta}{2})^2(\mathcal{L}_{\uparrow,\downarrow}(\Delta_p)\pm\Omega_z^2)},
 \end{eqnarray} 
\end{widetext}
with 
 \begin{eqnarray}
\mathcal{L}_{\uparrow,\downarrow}(\Delta_p)=&& 4[(i\Delta_p\mp\mathcal{M})^2+(\alpha-\frac{\delta}{2})^2].
 \end{eqnarray} 
The above expression for $c_{-}(\Delta_p)$ reveals the complete behavior of probe transmission under influence of all subsystems and shows how dressed states will imprint the spin-texture in probe field during transitions.

\subsection{Influence of atom-cavity coupling on cavity transmission}

\begin{figure}[h]
\includegraphics[width=9cm]{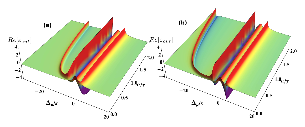}
\caption{The absorption $Re[\varepsilon_{out}]$ of the cavity probe transmission spectrum verses $\Delta_p/\kappa$ and $\alpha/\Omega$. (a) and (b) show $Re[\varepsilon_{out}]$ at $G=0.5\Omega$ and $2\Omega$, respectively. The remaining parameters are same as in Fig.2 of main text.}
\label{fig1}
\end{figure}
In order to uncover the effects of effective atom-field coupling on the probe transmission, we plot probe absorption $Re[\varepsilon_{out}]$ as a function of $\Delta_p/\kappa$ and $\alpha/\Omega$ for different atom-field couplings $G/\Omega$, as shown in Fig. \ref{fig1}.
 The position of the third structure initially moves away from the resonant point (i.e. $\Delta_p=0$) to the $\Delta_p<<0$ region, however, if we further increase $\alpha/\Omega$, it will saturates towards $\Delta_p/\Delta$, as shown in Fig. \ref{fig1}(a) and \ref{fig1}(b), where the atom-field coupling is $G=0.5\Omega$ and $G=2\Omega$, respectively. The increase in $G/\Omega$ increases the strength of transparency windows, because at stronger atom-field coupling, the quantum interference among the energy states becomes more robust causing strong photon-photon interactions \cite{Ref1,Ref2,Ref3}.                

\subsection{Inter-species interaction effects on transparencies}
\begin{figure}[h]
	\includegraphics[width=7cm]{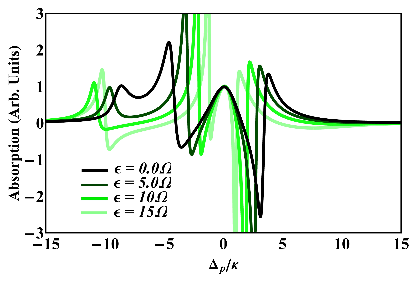}
	\caption{The absorption $Re[\varepsilon_{out}]$ verses $\Delta_p/\kappa$ under influence of inter-atomic interactions $\epsilon/\Omega$. The color configuration from darker to lighter illustrates the increase in $\epsilon/\Omega$. The remaining parameters are same as mentioned in Fig.2 of main text.}
	\label{fig11}
\end{figure}
 It will be very interesting to see how these photonic nonlinearities will behave when two atomic species will interact with each other. As the dressed states of atoms are acting as two separate oscillators coupled with each other via cavity field, therefore, their inter-interaction will definitely modify the probe field transitions. The inter-species interactions $\epsilon$ is appeared to further enhancing the effects of SO-coupling on probe transmission, as illustrated in Fig. \ref{fig11}. One can note the appearance of third EIT-like window at low $\alpha$, because of $\epsilon$, which previously appears at higher $\alpha$. These enhancements agree to the splitting of atomic dressed states with SO-coupling and increase in quantum interferences with their interactions.

\subsection{Raman coupling and detuning verses cavity transmission}
\begin{figure}[h]
	\includegraphics[width=9cm]{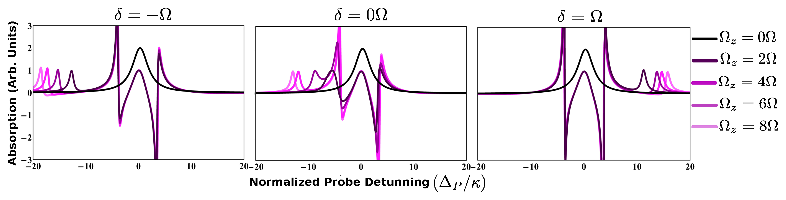}
	\caption{The absorption $Re[\varepsilon_{out}]$ verses $\Delta_p/\kappa$ for various values of applied Raman coupling $\Omega_z/\Omega$. (a), (b) and (c) show the absorption spectrum for $\delta=-\Omega$, $=0\Omega$ and $=\Omega$, respectively. The $G=0\Omega$ for the black curve and $G=10\Omega$ for all other curves. The SO-coupling is $\alpha=2\Omega$ while other parameters are same as in Fig.2 of main text.}
	\label{fig7}
\end{figure}

The Raman coupling $\Omega_z$, emerging from the Raman processes between the dressed states, is appeared to be directly proportional to the SO-coupling \cite{Ref34,Ref48} and its not possible to observe the effects of one while assuming other absent, specially in cavity systems. Therefore, the Raman coupling will similarly generate transparency windows via nonlinear optical interactions as SO-coupling is inducing, as illustrated in Fig. \ref{fig7}(b), where $\delta=0$ and $\alpha=2\Omega$ while $\Omega_z$ is being increased over the ratio of $\Omega$. One can note that, in the absence of $\Omega_z/\Omega$ and $G/\Omega$ (black curves), there is no transparency appearing in absorption, however, when we apply $G=10\Omega$ and increase the value of $\Omega_z/\Omega$, both atomic-cavity EIT and SO-couping induced transparencies start appearing, similar like SO-coupling. It reveals that the gap emerging in the eigen-energies of the atomic states with $\Omega_z$ \cite{Ref48} cause the new type of quantum interferences. As the strength of Zeeman gap appearing in atomic level states, $\hbar\omega_z$, significantly depends on the Raman detuning $\delta$, therefore, the value $\delta$ changes the position of third EIT-like window, as can be seen in Fig. \ref{fig7}(a) and \ref{fig7}(c), where $\delta=\Omega$ and $=-\Omega$, respectively. The symmetry of atomic eigen-energy spectrum breaks with the negative sign of $\delta$ to the negative of quasi-momentum and to the positive with positive $\delta$ \cite{Ref48}. Similarly, the transparency window moves towards left-side with negative $\delta$ (Fig. \ref{fig7}(a)) and right-side with positive $\delta$ (Fig. \ref{fig7}(c)). Thus, this new type of transparencies can also be labeled as gap induced transparencies which could be very significant for topological quantum computation. 

\subsection{The dynamics of fast and slow light}
\begin{figure*}[htp]
	\includegraphics[width=14cm]{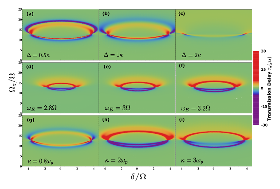}
	\caption{The slow and fast dynamics of probe light (with $\tau_g$) as a function of $\Omega_z/\Omega$ and $\delta/\Omega$ when $P=1.5P_p$. (a-c) illustrate the effects of system detuning $\Delta/\kappa$. (d-f) accommodate the influences of control field frequency $\omega_E/\Omega$. (g-i) show the influences of cavity decay rate $\kappa/\omega_p$. (g) is for gain regime ($\kappa/\gamma<1$) while (h) and (i) are for lossy regime ($\kappa/\gamma>1$). The remaining parameters used are the same as in Fig.2 of main text.}
	\label{fig9}
\end{figure*}
The transmission delay $\tau_g$ that is demonstrated under the influence of dressed states, as shown in Fig. 4(c) of main text, can be significantly altered by the other system parameters, as illustrated in Fig. \ref{fig9}. The effective cavity detuning $\Delta$ plays an important rule here because when we increase $\Delta$ it became off-resonant with atomic mode resulting in a weak cavity mode which eventually decreases the effects of atomic dressed states on probe transmission, as can be seen in Fig. \ref{fig9}(a-c). Both slow and fast probe transmission are decreased with an increase in $\Delta$ and follow complete circular pattern for $\Omega_z/\Omega$ and $\delta/\Omega$ with two maximum values at $\delta/\Omega$. However, by increasing the control field frequency $\omega_E$, one can enhance these dynamics over $\delta/\Omega$, shown in Fig. \ref{fig9}(d-f). The semi-circular dynamics of both fast as well as slow are appeared to shrunk and expanded with lower and higher values of control field frequency, respectively. It is obvious because at small control frequencies the probe field will be more sensitive to the Zeeman field effects, due to the decrease in frequency difference between pump and probe field. As we have already discussed the influence of loss ($\kappa/\gamma>1$) and gain ($\kappa/\gamma<1$) on probe transmission in main text, so it will be interesting to see the effects of dressed states on fast and slow probe transmission in different cavity decay rate $\kappa$ scenario, see Fig. \ref{fig9}(g-i). At low $\kappa$ ($\kappa/\gamma<1$), the dressed states interacts with probe in such away that the maximum value of both fast and low light follow complete circular pattern depending upon the $\Omega_z/\Omega$ and $\delta/\Omega$, similarly as in $\Delta$ case. But when we increase $\kappa$ ($\kappa/\gamma>1$), the $\tau_g$ moves to a single maxima (or semi-circular) domain of fast and slow probe light, as shown in main text. These findings show that the chosen system parameters provide a flexibility to switch the fast and slow dynamics of probe light.   
  
 \subsection{Fano Resonances with SO-coupling}
 
 \begin{figure}[h]
 	\includegraphics[width=9cm]{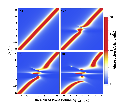}
 	\caption{(a) and (b) show absorption $Re[\varepsilon_{out}]$ verses $\Delta_p/\kappa$ and $\Delta/\kappa$ when $G=0\Omega$ and $10\Omega$, respectively, in absence of Raman coupling $\Omega_z=0$. (c) and (d) illustrate $Re[\varepsilon_{out}]$ for $\Omega_z=1\omega$ and $6\Omega$, respectively, when $G=10\Omega$. The SO-coupling $\alpha=2\Omega$ and $\delta=0$, while the remaining parameters are same as in Fig.2 of main text.}
 	\label{fig15}
 	\end{figure}
The Fano resonances occur in an EIT, or in any optical, system when the quantum interference produces an asymmetric line-shapes with a slight variation in resonant configuration and possess great importance in cavity-atom spectroscopy \cite{Ref50,Ref51}. In our system, because of SO- and Raman coupling mediated multiple transparencies, the optical (or Fano) interactions generate multiple asymmetric Fano resonances or line-shapes \cite{Ref28,Ref29,Ref50}, as illustrated in Fig. \ref{fig15}. At $G=0$, absorption spectrum show a linear line-shape without any transparency and resonance, see Fig. \ref{fig15}(a). However, in the case of just atoms, means no SO-coupling to produce $\Omega_z$, the transmission spectrum show a single break (or cut) in the absorption line-width which demonstrates the asymmetric Fano line-shapes, see Fig. \ref{fig15}(b). Further, when we apply SO-coupling, the strength of $\Omega_z/\Omega$ produces multiple breaks in the cavity absorption spectrum, as shown in Fig. \ref{fig15}(c) at $\Omega_z=\Omega$ and \ref{fig15}(d) at $\Omega_z=6\Omega$. The reason behind these interesting features is same as for transparencies. The occurrence of transparencies generates multiple dips in cavity transmission where the Fano interactions at non-resonant (or near resonant) system detuning $\Delta\ne0$ resulting in asymmetric line-shapes for Fano resonances. Thus, inclusion of SO-coupled atomic dressed states in cavity not only gives birth to new type of transparencies but also generates multiple Fano resonances for modern optical spectroscopy.